\documentclass{JHEP3}
\usepackage{amsmath}
\usepackage{amsthm}
\usepackage{amsfonts}
\usepackage{amssymb}

\def \Fig#1#2#3 {
\begin{figure}
\centering
\epsfxsize=#2cm \epsfbox{#1.eps}
\caption{#3}
\label{#1}
\end{figure}
}

\newcount\figno
\def\fig#1#2#3{
\par\begingroup\parindent=0pt\leftskip=1cm\rightskip=1cm\parindent=0pt
\baselineskip=15pt
\global\advance\figno by 1
\epsfxsize=#3
\centerline{\epsfbox{#2}}
\vskip 12pt
{\bf \small Figure \the\figno:} {\small #1}\par
\endgroup\par
}
\def\figlabel#1{\xdef#1{\the\figno
\mbox{ }}}
\def\encadremath#1{\vbox{\hrule\hbox{\vrule\kern8pt\vbox{\kern8pt
\hbox{$\displaystyle #1$}\kern8pt}
\kern8pt\vrule}\hrule}}

\title{ WZW models as  mutual super Poisson-Lie T-dual sigma models}
\author{\\
Department of Physics, Faculty of science,
Azarbaijan Shahid Madani University, 53714-161, Tabriz, Iran\\
E-mail: \email{a.eghbali@azaruniv.edu}, \email{rezaei-a@azaruniv.edu}
}

\keywords{ Conformal and W Symmetry, String Duality, Sigma Models}

\abstract{A WZW model on the Lie supergroup $(C^3+A)$ is constructed. It is shown that this model contains super Poisson-Lie symmetry with
the dual Lie supergroup ${C^3  \oplus A_{1,1}}_|.i$. Furthermore,
we show that the dual model is also equivalent to the WZW model on isomorphic Lie supergroup $(C^3+A).i$. In this manner,
because of  isomorphism of the $({\cal C}^3+{\cal A})$ with a Manin supertriple, it is shown that the N=(2,2) structure is preserved under the
super Poisson-Lie T-duality transformation.}

\begin{document}





\section{Introduction}

The WZW models on Lie supergroups  have recently received
considerable attention, because of their relations to local
logarithmic conformal field theories
\cite{schom}-\cite{Saleur}. Superstring theory on $AdS$
backgrounds relates to WZW model on the Lie supergroups and the
coset superspaces  \cite{Henneaun}-\cite{Bershadsky}. On the other hand, duality symmetries play an
important role in string theory. They are specific to string
theory and their study led to important insights in understanding
the geometry of space-time from the string point of view. A very
important symmetry of string theories, or more generally,
two-dimensional sigma models is the T-duality \cite{Busher}.
It has been shown that the conformal symmetry is preserved under the Abelian duality \cite{Verlinde}, and also this duality has been investigated
in the WZW model (see for example \cite{Kiritsis}). Furthermore for the non-Abelian duality case \cite{Ossa}, it has been shown that the conformal
symmetry is preserved when the trace of the adjoint representation of the isometry group is zero \cite{Givoen}.
The Poisson-Lie T-duality \cite{K.S1} is a generalization of Abelian
and non-Abelian  target space duality
(T-duality). This generalized duality is associated with the two
groups forming a Drinfel'd double such that the duality
transformation exchanges their roles. Up to now, there is one
example for the conformal Poisson-Lie T-dual sigma models
\cite{K.S3}; such that the duality relates the $SL(2,R)$ WZW
model to a constrained sigma model. Recently the generalization
of the Poisson-Lie T-duality in sigma models on supermanifolds
has been performed in \cite{{ER}, {cosmology}}. Furthermore, we have
shown  that the WZW model on the Lie supergroup $GL(1|1)$  has
also super Poisson-Lie symmetry when the dual Lie supergroup is
${B \oplus A \oplus A_{1,1}}_|.i$ \cite{ER7}. Here we first show
that the WZW model on the Lie supergroup $(C^3+A)$ contains  super
Poisson-Lie symmetry, then we describe a pair of super
Poisson-Lie T-dual sigma models which is associated with the
$\left((C^3+A)\;,\;{C^3 \oplus {A}_{1,1}}_|.i\right)$  Drinfel'd
superdouble such that the original model is the WZW model on the
$(C^3+A)$; moreover, we show that the dual model is also equivalent to the WZW model on isomorphic Lie supergroup $(C^3+A).i$.
Thus, in this way we have found an example in which
conformal invariance is preserved under the super Poisson-Lie T-duality
transformation. In fact,  this is the first example of the
conformal  sigma models related by super Poisson-Lie T-duality
which has  so far been constructed. Furthermore,  because of isomorphism of the $({\cal C}^3+{\cal A})$ with a Manin supertriple  we show that
the N=(2,2) structure is preserved under the
super Poisson-Lie T-duality transformation for this example.

The structure of the paper is as follows. In Section 2, for  introducing of the notations and self-containing of the subject, we briefly review
some aspects of the super Poisson-Lie symmetry and T-dual sigma model on Lie
supergroup \cite{ER}. In Section 3, by using
the condition of the super Poisson-Lie symmetry  we show that the WZW
model on $(C^3+A)$ Lie supergroup contains super Poisson-Lie symmetry
when the dual Lie supergroup is ${C^3  \oplus
 A_{1,1}}_|.i$. Then, we obtain the mutually T-dual sigma models on the Drinfel'd
superdouble  $\left((C^3+A)\;,\;{C^3 \oplus
{A}_{1,1}}_|.i\right)$ in Section 4 such that the original model
is the same as the $(C^3+A)$ WZW model. The proof of the
conformal invariance for the dual model is also given in Section
4. Furthermore, we show that T-dual sigma models have flat backgrounds. Indeed, the dual model is equivalent to the WZW model on isomorphic Lie supergroup $(C^3+A).i$. Finally, in Section 5 by use of the fact that the algebraic representation of the N=(2,2) structure is a Manin triple and
the $({\cal C}^3+{\cal A})$ Lie superalgebra is a Manin supertriple; we show that the N=(2,2) structure is preserved under the
super Poisson-Lie T-duality transformation.

\section{A review of super Poisson-Lie T-duality}

Consider a nonlinear sigma model described by a supersymmetric
metric $G_{\Upsilon\Lambda}$ and a super antisymmetric two-form
field $B_{\Upsilon\Lambda}$ on the target  supermanifold $M$ (the
superdimension of the supermanifold is written as ($\#$ bosons $|$
$\#$ fermions)=$(d_B|d_F)$, which,  because of the invertibility of the
metric $G_{\Upsilon\Lambda}$, $d_F$ must be even \cite{D}) with the action
\footnote{ We note that the $(-1)^{^{|\Upsilon|}}$ denotes the parity of
$\Upsilon$ where  $|\Upsilon|=0$ for the bosonic coordinates and $|\Upsilon|=1$ for the fermionic
coordinates.   Here and in the following we use the notation
presented by DeWitt  \cite{D}, e.g.,
$(-1)^{\Upsilon}:=(-1)^{|\Upsilon|}$.}
$$
S=\frac{1}{2}\int\!d\xi^{+} d\xi^{-}\;(-1)^{^{|\Upsilon|}}   \partial_{+}\Phi
^{^{\Upsilon}}\;{{\cal E}}_{\Upsilon\Lambda} \partial_{-}\Phi^{^{\Lambda}}~~~~~~~~~~~~~~~~~~~~
$$
\vspace{-3mm}
\begin{equation}\label{A.1}
=\frac{1}{2}\int\!d\xi^{+} d\xi^{-}(-1)^{^{|\Upsilon|}}   \partial_{+}\Phi
^{^{\Upsilon}}\; (G_{\Upsilon\Lambda}
+B_{\Upsilon\Lambda})\;\partial_{-}\Phi^{^{\Lambda}},~~~
\end{equation}
where $\xi^{\pm} \;\equiv\;
\frac{1}{2}(\tau \pm \sigma)$ are the standard light-cone variables on the worldsheet and $\Phi{^{^ \Upsilon}}$ are coordinates on $M$, which
include the bosonic coordinates $X^{^{\mu}} (\mu=0,\cdots,d_B-1)$  and the fermionic
ones $\Theta ^{^{\alpha}} (\alpha=1,\cdots,d_F)$, and the labels $\scriptsize{\Upsilon}$
and $\scriptsize {\Lambda}$ run over $(\mu , \alpha)$. Let now
\begin{equation}\label{A.2}
\star {J_i}\;
=\;(-1)^{\Lambda +\Upsilon\Lambda}\;{{V_i}^{}}^{^{\Lambda}}
\;{\cal E}_{_{{\Upsilon\Lambda
}}}\;\partial_{+} \Phi ^{^{\Upsilon}}\;d\xi^{+} -(-1)^{\Lambda}\;{{V_i}^{}}^{^{\Lambda}}\;{\cal
E}_{_{{\Lambda\Upsilon}}}\;\partial_{-} \Phi^{^{\Upsilon}} d\xi^{-},
\end{equation}
be the Hodge star of Noether's current one-forms corresponding to
the right action of the Lie supergroup $G$ on the target $M$, in which
${{V_i}}^{^{\Upsilon}}(\Phi)$'s are the left invariant supervector
fields defined with left derivative \cite{ER}. If  the  forms
$\star {J_i}$ not be closed and on the extremal surfaces
satisfy  the Maurer-Cartan equation \cite{D}, we say that the
sigma model has the super Poisson-Lie symmetry with respect to
the Lie supergroup $\tilde G$ (the dual Lie supergroup to ${ G}$)
\cite{ER}. It is a condition on ${\cal E}_{_{{\Upsilon\Lambda
}}}$ so that it can be formulated as \cite{ER}
\begin{equation}\label{A.3}
{\cal L}_{V_i}({\cal E}_{_{{\Upsilon\Lambda }}})
=(-1)^{i(\Upsilon+k)}\;
 {\cal E}_{_{{\Upsilon\Xi
}}} {(V^{st})^{^{\Xi}}}_{k}\;(\tilde
{\cal{Y}}_i)^{kj}\;{{_jV}}^{^{\Omega}}\;{\hspace{-0.5mm}_{_{\Omega}}
{ {\cal{E}}}}\hspace{-0.5mm}_{_{ \Lambda}},
\end{equation}
where $(\tilde {\cal{Y}}_i)^{jk} = -{{\tilde f}_{\;\;\;\;i}}^{jk}$
are the adjoint representations of Lie superalgebra ${\tilde {\mathcal G}}$ (the dual
Lie superalgebra to $\mathcal{ G}$)
and ``$st$'' stands for the {\it supertranspose} \cite{D}.

Let us suppose now that $G$ acts transitively and freely on $M$; then
the mutually $T$-dual sigma models have target spaces in the Lie supergroups $G$ and $\tilde G$ and are defined
by the actions\footnote{The construction of the models has been described in \cite{ER}.}
\begin{equation}\label{A.4}
S\;=\;\frac{1}{2}\int\!d\xi^{+}
d\xi^{-}\;(-1)^i {R_{+}}{\hspace{-1mm}^{(l)^i}}\;  {E_{ij}^+}(g)\;{R_{-}}^{{\hspace{-1mm}(l)^j}},~~~~~~~~g\in G,
\end{equation}
\begin{equation}\label{A.5}
\tilde S\;=\;\frac{1}{2}\int\ d\xi^{+} d\xi^{-}\;(-1)^j {{\tilde
R^{(l)}_{+}}\hspace{-0.5mm}_i}\; {{{{\tilde E}^{+ij}}}}(\tilde g)
\;{{\tilde R}_{-j}}^{{\hspace{0mm}(l)}},~~~~~~~~~{\tilde g}\in {\tilde G},
\end{equation}
where
\begin{equation}\label{A.5.1}
{R}{\hspace{0mm}^{(l)^i}}=(dg g^{-1})^i=(-1)^{^{\Upsilon}} d\Phi^{^{\Upsilon}} \;
 R_{_{\Upsilon}}^{(l)^{~i}},\hspace{3cm}
\end{equation}
and
\begin{equation}\label{A.5.2}
{{\tilde R^{(l)}}\hspace{-0.5mm}_i}=(d \tilde g {\tilde
g}^{-1})_i=(-1)^{^{\Upsilon}} d {\tilde \Phi}^{^{\Upsilon}}
\;{{\tilde  R^{(l)}}\hspace{-0.5mm}_{{\Upsilon} i}},\hspace{3cm}
\end{equation}
are right invariant one-forms (defined with left derivative).
Here the matrices $E(g)$ and ${\tilde E}(\tilde g)$ are defined
in the following way\footnote{Here, one must use of
superdeterminant and superinverse formulas \cite{D}.}
\begin{equation}\label{A.6}
{ E^{+}(g)}\;=\;\Big(\Pi(g)+ ({{E_0}^{\hspace{-1mm}+}})^{-1}(e)\Big)^{-1},~~~~~~~~\tilde {E^{+}}(\tilde g)\;=\;\Big(\tilde \Pi(\tilde g)+ {({{\tilde
E_0}^{+}})^{-1}(\tilde e) \Big)}^{-1},
\end{equation}
in which ${{E_0}^{\hspace{-1mm}+}}(e)$ and ${{\tilde
E_0}^{+}}(\tilde e)$ are constant matrices such that in the vicinity of the origin of the Lie supergroup ($g=e$ and $\tilde g= \tilde e$) are related to each
other as
\begin{equation}\label{A.7}
{_i{E_{0_{j}}}}^{\hspace{-2mm}+}(e) {{\tilde E_0}}^{+^{jk}}(\tilde e)\;=\;{_i \delta}^{~k},
\end{equation}
and $\Pi^{ij}(g)=(-1)^k b^{ik}(g)\; {(a^{-1})_k}^j(g)$, where the matrices $a(g)$ and $b(g)$ are constructed in the following way
$$
g^{-1} X_i\; g\; =\;(-1)^j\;a_i^{\;\;j}(g)\; X_j,~~~~~~~~~
$$
\begin{equation}\label{A.9}
~~~~~~g^{-1} \tilde{X}^i g\; =\; (-1)^j\;{b^{ij}(g)}\;{X_j} +
d^i_{\;\;j}(g)\; \tilde{X}^j.
\end{equation}
Here $X_i$ and $\tilde X^i$ are basis of the respective Lie
superalgebras $ {\bf \mathcal{G}}$ and $\tilde {\bf \mathcal{G}}$
where form a pair of maximally superisotropic Lie subsuperalgebras
into the Drinfel'd superdouble \footnote {A Drinfel'd superdouble
\cite{ER4} is a Lie superalgebra ${\cal{D}}$ which decomposes
into the direct sum, as supervector spaces, of two maximally
superisotropic Lie subsuperalgebras ${\mathcal{G}}$ and $\tilde
{\mathcal{G}}$, each corresponding to a Poisson-Lie supergroup
($G$ and $\tilde G$), such that the Lie subsuperalgebras are duals of
each other in the usual sense, i.e., $\tilde
{\mathcal{G}}={\mathcal{G}}^*$.} so that
$$
< X_i , X_j >\; =\; < {\tilde X}^i , {\tilde X}^j
>\;=\;0,~~~~~~~~~~~~~~~
$$
\vspace{-5mm}
\begin{equation}\label{A.10}
~~< X_i , {\tilde X}^j >\; =\;  (-1)^{ij} < {\tilde X}^j , X_i
>\;=\;(-1)^{i}{_i\delta} \hspace{1mm}^j,
\end{equation}
and their structure constants satisfy in the following mixed super Jacobi identities \cite{ER1}:
\begin{equation}\label{A.10.1}
{f^m}_{jk}{\tilde{f}^{il}}_{\; \; \; \; m}=
{f^i}_{mk}{\tilde{f}^{ml}}_{\; \; \; \; \; j} +
{f^l}_{jm}{\tilde{f}^{im}}_{\; \; \; \; \; k}+ (-1)^{jl}
{f^i}_{jm}{\tilde{f}^{ml}}_{\; \; \; \; \; k}+ (-1)^{ik}
{f^l}_{mk}{\tilde{f}^{im}}_{\; \; \; \; \; j}.
\end{equation}

\section{Super Poisson-Lie symmetry in the $(C^3 + A)$  WZW model}

In what follows we shall construct a WZW model based on  the Lie
supergroup $(C^3 + A)$ of dimension-$(2|2)$. Indeed in \cite{ER4}
it has been shown that this Lie supergroup is the same as the
Drinfel'd superdouble $D=\left((A_{1,1} + A) ~,~I_{(1|1)}
\right)$ in such a way  that its Lie superalgebra $({\cal C}^3 +{\cal A})$
is a $(1|1)$-dimensional Lie superbialgebra \cite{B} and has the
following explicit description:
\begin{equation}\label{L.1}
[H , Q_-]=Q_+,~~~[H , Q_+]=0,~~~\{Q_- , Q_-\}=Z,~~~\{Q_+ ,
Q_{\pm}\}=0,~~~[Z~ ,~ .]=0,
\end{equation}
where $H, Z$ and $Q_+, Q_-$ are bosonic generators and fermionic ones, respectively. To
define a WZW model, one needs a bilinear form $\Omega_{ij}$ in the generators $X_i$, which is
supersymmetric $(\Omega_{ij}=(-1)^{ij} \Omega_{ji})$  ad-invariant metric on ${\mathcal G}$ and satisfies
in the following relation
\begin{equation}\label{L.2}
f_{ij}^{\;\;l} \;\Omega_{lk}+(-1)^{jk} f_{ik}^{\;\;l} \;\Omega_{lj}\;=\;0.
\end{equation}
A non-degenerate solution to (\ref{L.2}) exists and is given by
\begin{equation}\label{L.3}
\Omega_{ij}=\left( \begin{tabular}{cccc}
              $b$&  $a$ & 0 & $c$\\
              $a$ & 0 & 0& $d$ \\
              $0$ & 0 & 0& $-a$ \\
              $c$ & $d$ & $a$ &0\\
                \end{tabular} \right),
\end{equation}
where $a, b, c$ and $d$ are real constants and $a$ is non-zero.
Then the WZW action on a Lie supergroup $G$ is given by \cite{ER7}
$$
S_{WZW}(g) =  \frac{k}{4\pi} \int_{\Sigma} d^2\xi\; (-1)^{i}
L^{\hspace{-0.5mm}(l)i}_{+}\;{\Omega}_{ij}\;
L^{\hspace{-0.5mm}(l)j}_{-}\hspace{7cm}
$$
\vspace{-2mm}
\begin{equation}\label{L.4}
\hspace{0cm}-~\frac{k}{24\pi} \int_{B} d^3 \xi (-1)^{i+jk}
\varepsilon^{ \gamma \alpha \beta}
L^{\hspace{-0.5mm}(l)i}_{\gamma}
\;{\Omega}_{il}\;L^{\hspace{-0.5mm}(l)j}_{\alpha}
({\cal{Y}}^l)_{jk}\; L^{\hspace{-0.5mm}(l)k}_{\beta},
\end{equation}
where the $L^{\hspace{-0.5mm}(l)i}_{\alpha}$'s are defined via $g^{-1}
\partial_{\alpha}g\;=\;(-1)^i L^{\hspace{-0.5mm}(l)i}_{\alpha} X_{i}$ (with $g:  \Sigma \rightarrow G$) and {\small $B$} is a three-manifold bounded by worldsheet $\Sigma$.
To write explicit form of the action (\ref{L.4}) on $G$ we choose  a general element of the Lie supergroup $(C^3 +A)$ as
\begin{equation}\label{L.5}
g\;=\; e^{\chi Q_-}  e^{y H+x Z}  e^{\psi Q_+},
\end{equation}
where $x(\tau , \sigma)$ and $y(\tau , \sigma)$ are bosonic fields while $\psi(\tau , \sigma)$ and $\chi(\tau , \sigma)$ are fermoinic ones.
Then $L^{\hspace{-0.5mm}(l)i}_{\alpha}$'s are found to be
$$
L^{\hspace{-0.5mm}(l)H}_{\alpha}\;=\;{\partial}_{\alpha}{y},~~~~~~~~~~~~
$$
\vspace{-3mm}
$$
L^{\hspace{-0.5mm}(l)Z}_{\alpha}\;=\;{\partial}_{\alpha}{x}-{\partial}_{\alpha}{\chi}~\frac{\chi}{2},
$$
\vspace{-3mm}
$$
L^{\hspace{-0.5mm}(l)Q_+}_{\alpha}\;=\;-{\partial}_{\alpha}{\psi}+{\partial}_{\alpha}{\chi} ~y,
$$
\vspace{-3mm}
\begin{equation}\label{L.6}
L^{\hspace{-0.5mm}(l)Q_-}_{\alpha}\;=\;-{\partial}_{\alpha}{\chi}.~~~~~~~~~~~~
\end{equation}
Furthermore, the terms that are being integrated over in (\ref{L.4}) are
calculated to be
$$
L^{\hspace{-0.5mm}(l)i}_{+}\;{_i\Omega}_j\;
L^{\hspace{-0.5mm}(l)j}_{-} =a[\partial_{+} y
\partial_{-} x + \partial_{+} x
\partial_{-} y -\partial_{+} \psi\;
\partial_{-} \chi + \partial_{+} \chi\;
\partial_{-} \psi~~~~~~~~~~~~~~~~~~
$$
\vspace{-4mm}
\begin{equation}\label{L.7}
~~~~~~~~~~~~~~~~~~~~~~~~~~~~~~~~+\partial_{+} y~\frac{\chi}{2}~
\partial_{-} \chi-\partial_{+} \chi ~ \frac{\chi}{2}~
\partial_{-}y],~~~~~~~~~
\end{equation}
{\small\begin{equation}\label{L.8} (-1)^{jk}
L^{\hspace{-0.5mm}(l)i}_{\gamma}
\;{_i\Omega}_l\;L^{\hspace{-0.5mm}(l)j}_{\alpha}
({\cal{Y}}^l)_{jk}\; L^{\hspace{-0.5mm}(l)k}_{\beta}\;=\;a
\partial_{\gamma} [y\partial_{\alpha} \chi
\partial_{\beta} \chi - \chi\;\partial_{\alpha} y \partial_{\beta}
\chi+\chi \partial_{\alpha} \chi \partial_{\beta} y].
\end{equation}}
Hence, the WZW action on the Lie supergroup $(C^3 +A)$ is written
as {\small\begin{equation}\label{L.9}
S_{WZW}(g) =  \frac{1}{2}
\int_{\Sigma} d\xi^{+} d\xi^{-}\;(\partial_{+} y
\partial_{-} x + \partial_{+} x
\partial_{-} y - \partial_{+} \psi\;
\partial_{-} \chi + \partial_{+} \chi\;
\partial_{-} \psi +  \partial_{+} y ~{\chi}~  \partial_{-} \chi ).
\end{equation}}
Here we have assumed that $b=c=d=0$ in (\ref{L.3}), and $a$ has
rescaled to $\frac{2\pi}{k}$. By identifying the action
(\ref{L.9}) with the sigma model of the form (\ref{A.1}), one can
read off the background matrix as
\begin{equation}\label{L.10}
{\cal E}_{\Upsilon\Lambda}=\left( \begin{tabular}{cccc}
              $0$&  $1$ & 0 & $\chi$\\
              $1$ & 0 & 0& $0$ \\
              $0$ & 0 & 0& $1$ \\
              $0$ & $0$ & $-1$ &0\\
                \end{tabular} \right).\hspace{6cm}
\end{equation}
As showned in the above, the background matrix is a function of
the fermionic field $\chi$ only. In order to find super
Poisson-Lie symmetry in  WZW model on the Lie supergroup
$(C^3+A)$, we need the left invariant supervector fields (defined
with left derivative) on the $(C^3+A)$.  By applying relations
(\ref{L.6}) in the relation
${{_iV}}^{^{\Upsilon}}=({{_{\Upsilon}}{L}}^{i})^{-1}$ \cite{ER}
the supervector fields are calculated to be
\begin{align}
{_{_H}V} &=\ \frac{ \overrightarrow{\partial}}{\partial y},\nonumber \\[2mm]
{_{_Z}V} &=\ \frac{ \overrightarrow{\partial}}{\partial x},\nonumber \\[2mm]
{_{_{Q_+}}V} &=\ -\frac{ \overrightarrow{\partial}}{\partial
\psi},\nonumber \\[2mm]
{_{_{Q_-}}V} &=\ -\frac{\chi}{2} \frac{
\overrightarrow{\partial}}{\partial  x}-{y}\frac{
\overrightarrow{\partial}}{\partial \psi}-\frac{
\overrightarrow{\partial}}{\partial \chi}.\label{L.11}~
\end{align}
Now by calculation of the Lie superderivatie of the bi-tensor
${\cal E}_{\Upsilon\Lambda}$ (\ref{L.10})  with respect to ${_{_{i}}V}$'s and
then by substituting  relations (\ref{L.10}) and (\ref{L.11}) on
the right hand side of Eq. (\ref{A.3}), after some calculations the non-zero structure
constants of the dual pair to the $({\cal C}^3 +{\cal A})$ Lie
superalgebra are found to be
\begin{equation}\label{L.12}
[\tilde Z ,\tilde Q_+]=-\frac{1}{2}\tilde
Q_-.~~~~~~~~~~~~\hspace{5cm}
\end{equation}
On the other hand we have been recently  classified all $(2|2)$-dimensional
decomposable Lie superalgebras \cite{ER6}. The Lie superalgebra
obtained in (\ref{L.12}) is isomorphic to the decomposable
trivial Lie superalgebra ${\cal C}^3 \oplus {\cal A}_{1,1}$
\cite{ER6} under the following transformation:
\begin{align}
\tilde H &=\ a Z',\nonumber \\[2mm]
\tilde Z & =\ b H' +c Z',\nonumber \\[2mm]
{\tilde Q}_+ & =\ -d {Q'}_+ -e {Q'}_-,\nonumber \\[2mm]
{\tilde Q}_- & =\ 2be {Q'}_+,\label{L.13}~
\end{align}
where $a, b, e \in \Re-\{0\}; c, d\in \Re$.  The generetors
$\{H', Z'\} $   and $\{Q'_+, Q'_-\}$ are bosonic  and fermionic
generators of the  Lie superalgebra ${\cal C}^3 \oplus {\cal
A}_{1,1}$, respectively. Hence we denote the dual pair to the
$({\cal C}^3 + {\cal A})$ by ${{\cal C}^3 \oplus {\cal
A}_{1,1}}_|.i$ with the same of the commutation relation obtained
in (\ref{L.12}). Two sets of generators  (\ref{L.1}) and
(\ref{L.12}) are dual to each other in the sense of (\ref{A.10}).
Nevertheless, one can check that  the Lie superbialgebra
$\left(({\cal C}^3 +{\cal A})\;,\;{{\cal C}^3 \oplus {\cal
A}_{1,1}}_|.i\right)$ satisfies mixed super Jacobi identities (\ref{A.10.1})
\cite{ER1}. The fact that the WZW model on the Lie supergroup
$(C^3+A)$ has super Poisson-Lie symmetry is quite interesting. In
the following, we shall construct the mutually  T-dual sigma
models on the Drinfel'd superdouble $\left((C^3+A)\;,\;{C^3
\oplus {A}_{1,1}}_|.i\right)$ and will show that the original
sigma model on the $(C^3 + A)$ is identical to
 WZW model on the $(C^3 + A)$.  Moreover, to prove the conformal invariance of
the dual model we will look at the one-loop $\beta$-function
equations and then will discuss that it is conformal invariant
for all loops.

\section{The $(2|2)$-dimensional super Poisson-Lie T-dual sigma models on the $\left((C^3+A)\;,\;{C^3 \oplus {A}_{1,1}}_|.i\right)$ }

To construct the mutually T-dual sigma models we need the
Derinfel'd superdouble which is simply a Lie supergroup $D$. The
decomposition of the superdouble into the pair of maximally
isotropic Lie subsuperalgebras ${\cal D}= {\mathcal G} \oplus {\tilde
{\mathcal  G}}$ is referred  to as the Manin supertriple $({\cal
D} , {\mathcal G} , {\tilde {\mathcal G}})$. Consider the Lie
superalgebra ${\mathcal G}=({\cal C}^3 + {\cal A})$ defined by
(anti)commutation relations (\ref{L.1}) and its dual ${\tilde
{\mathcal G}}={{\cal C}^3 \oplus {\cal A}_{1,1}}_|.i$ which is
generated by relation (\ref{L.12}). The Lie superalgebra of the
Derinfel'd superdouble which we refer to as the $\left(({\cal
C}^3 +{\cal A})\;,\;{{\cal C}^3 \oplus {\cal
A}_{1,1}}_|.i\right)$ is denoted by generators $\{ H, Z, \tilde
H, \tilde Z; Q_+, Q_-, \tilde Q_+, \tilde Q_-\}$ and the
following non-zero (anti) commutation relations
$$
[H , Q_-]=Q_+,~~~~~~~~\{Q_- , Q_-\}=Z,~~~~~~~[\tilde Z , \tilde
Q_+ ]=-\frac{1}{2}\tilde Q_-,~~~~~~~~~~~~~~~~~~~~~~~~
$$
\vspace{-4mm}
\begin{equation}\label{p.1}
~[H , \tilde Q_+]=-\tilde Q_-,~~~~~~~~[ Q_- , \tilde
Z]=-\frac{1}{2} Q_+ - \tilde Q_-,~~~~\{Q_- , \tilde
Q_+\}=-\frac{1}{2} Z-\tilde H.
\end{equation}
In order to construct the  corresponding original sigma model with
the Lie supergroup $(C^3+A)$ as the target space, we use the same
 parametrization (\ref{L.5}). Inserting this parametrization in
(\ref{A.5.1}) we find
\begin{equation}\label{P.2}
\partial_{\pm} g ~{ g}^{-1}\;=\; {\partial_{\pm}}{ y}~ H+ ({\partial_{\pm}}{ x}+ {\partial_{\pm}}{\chi} \frac{\chi}{2})~ Z
+ ({\partial_{\pm}}{\psi}-{\partial_{\pm}}{ y} \chi)~ Q_+ +
{\partial_{\pm}}{\chi}~ Q_-,
\end{equation}
from which we can obtain ${{R}^{(l)}_\pm}^i$'s as
$$
{{R}^{(l)}_\pm}^H\;=\;{\partial_\pm}y,~~~~~~~~~~~~~~~~~~~~~~~~
$$
$$
{{R}^{(l)}_\pm}^Z\;=\;{\partial_\pm}x+{\partial_\pm}\chi
\frac{\chi}{2},~ ~~~~~~~~~~~
$$
$$
{{R}^{(l)}_\pm}^{Q_+}\;=\;{\partial_\pm}y \chi-
{\partial_\pm}\psi,~~~~~~~~~~~~~~
$$
\begin{equation}\label{P.3}
{{R}^{(l)}_\pm}^{Q_-}\;=\;-{\partial_\pm}\chi.~~~~~~~~~~~~~~~~~~~~~~~
\end{equation}
Then the matrices $a(g),~ b(g)$ and $d(g)$ from Eq. (\ref{A.9})
read
$$
a_i^{~j}(g)=\left(\begin{tabular}{cccc}
              $1$ & $0$ & $-\chi$ & $0$\\
              $0$ & $1$ & $0$ & $0$ \\
              $0$ & $0$ & $-1$ & $0$ \\
              $0$ & $-\chi$ & $y$ & $-1$\\
                \end{tabular} \right),~~~b^{ij}(g)=\left( \begin{tabular}{cccc}
              $0$ & $0$ & $0$ & $0$\\
              $0$ & $0$ & $-\frac{\chi}{2}$& $0$ \\
              $0$ & $\frac{\chi}{2}$ & 0& $0$ \\
              $0$ & $0$ & $0$ & $0$\\
                \end{tabular} \right),
$$
\vspace{-2mm}
\begin{equation}\label{P.4}
d^i_{~j}(g)=\left( \begin{tabular}{cccc}
              $1$    & $0$ & $0$ & $0$\\
              $0$    & $1$ & $0$ & $\chi$ \\
              $\chi$ & $0$ & $1$ & $y$ \\
              $0$    & $0$ & $0$ & $1$\\
                \end{tabular} \right).~~~~~~~~~~~~~~~~~~~~~~~~~~~~~~~~~~~~~~~~~~
\end{equation}
Here we choose the sigma model matrix $E_0^+(e)$ at the unit element
of $(C^3+A)$ as
\begin{equation}\label{P.5}
E_{0_{ij}}^+(e)=\left( \begin{tabular}{cccc}
              $0$ & $1$ & $0$ & $0$\\
              $1$ & $0$ & $0$ & $0$ \\
              $0$ & $0$ & $0$ & $1$ \\
              $0$ & $0$ & $-1$ & $0$\\
                \end{tabular} \right).
\end{equation}
By a direct application of the first formula of (\ref{A.6}), the
corresponding sigma model with the  Lie supergroup $(C^3+A)$  is
worked out as follows:
{\small\begin{equation}\label{P.6}
S\;=\;\frac{1}{2} \int d\xi^{+} d\xi^{-}\;(\partial_{+} y
\partial_{-} x + \partial_{+} x
\partial_{-} y - \partial_{+} \psi\;
\partial_{-} \chi + \partial_{+} \chi\;
\partial_{-} \psi +  \partial_{+} y ~{\chi}~  \partial_{-} \chi ).
\end{equation}}
The above action is nothing but the $(C^3+A)$ WZW action. We have
thus showed that:

\smallskip
\smallskip
 ~{\it The super Poisson-Lie duality relates the
$(C^3+A)$ WZW model to a sigma model defined on the Lie
supergroup $(C^3+A)$ when the dual Lie supergroup is ${C^3 \oplus
{A}_{1,1}}_|.i$.}\\

Let us now turn into the dual sigma model.  We parametrize the
corresponding Lie supergroup  ${C^3 \oplus
{A}_{1,1}}_|.i$ with bosonic coordinates $\{\tilde x, \tilde y\}$ and fermionic ones  $\{\tilde \psi, \tilde \chi\}$ so that its elements can
be written in the same parametrization (\ref{L.5}) as:
\begin{equation}\label{P.7}
\tilde g\;=\; e^{\tilde \chi \tilde Q_-}  e^{\tilde y \tilde H+ \tilde x \tilde Z}  e^{\tilde \psi \tilde Q_+},~~~~~~~~~~~~~~~~~~~~~~~
\end{equation}
from which we can read off the ${{{\tilde R}_{+i}}^{(l)}}$'s as
\begin{equation}\label{P.8}
{\tilde R}_{\pm_{\tilde H}}^{(l)}= {\partial_\pm}{\tilde y},~~~~ {\tilde R}_{\pm_{\tilde Z}}^{(l)}= {\partial_\pm}{\tilde x},~~~ ~
{\tilde R}_{\pm_{\tilde Q_+}}^{(l)}= {\partial_\pm}{\tilde \psi},~~~~ {\tilde R}_{\pm_{\tilde Q_-}}^{(l)}= -{\partial_\pm}{\tilde \psi} \frac{\tilde x}{2}
+{\partial_\pm}{\tilde \chi},
\end{equation}
and then using Eqs. (\ref{A.9}) for the dual model we obtain
\begin{equation}\label{P.9}
{\tilde \Pi}_{ij}(\tilde g) \;=\;\left( \begin{tabular}{cccc}
              $0$ & $0$ & $0$ & $-{\tilde \psi}$\\
              $0$ & $0$ & $0$ & $0$ \\
              $0$ & $0$ & $0$ & $0$ \\
              ${\tilde \psi}$ & $0$ & $0$ & $-{\tilde x}$\\
                \end{tabular} \right).
\end{equation}
Finally, by substituting the above relation in the second equation of (\ref{A.6}) the dual model action is obtained as follows:
$$
\tilde S \; = \;  \frac{1}{2} \int  d\xi^{+}
d\xi^{-}\; ( \partial_{+} {\tilde y}  \partial_{-} {\tilde x}+  \partial_{+} {\tilde x}  \partial_{-} {\tilde y}
+\partial_{+} {\tilde x}~ \tilde \psi~ \partial_{-}{
\tilde \psi}+\partial_{+}  {\tilde \psi} \;{\tilde \psi}
\partial_{-} {\tilde x}
$$
\vspace{-5mm}
\begin{equation}\label{P.10}
~~~~~~~~~~~~~~~~~~~~~- \partial_{+}  {\tilde \psi}~{\tilde x}~ \partial_{-} {\tilde \psi}-\partial_{+} {\tilde \psi}~\partial_{-}{ \tilde \chi}+ \partial_{+}{
\tilde \chi} \;\partial_{-} {\tilde \psi}).
\end{equation}
The above action can be rewritten as the following one up to total derivatives
{\small \begin{equation}\label{P.10.1}
\tilde S \; = \;  \frac{1}{2} \int  d\xi^{+}
d\xi^{-}\; ( \partial_{+} {\tilde y}  \partial_{-} {\tilde x}+  \partial_{+} {\tilde x}  \partial_{-} {\tilde y}
-\partial_{+} {\tilde \psi}~\partial_{-}{ \tilde \chi}+ \partial_{+}{
\tilde \chi} \;\partial_{-} {\tilde \psi}- \partial_{+}  {\tilde \psi}~3{\tilde x}~ \partial_{-} {\tilde \psi}).
\end{equation}}
Comparing  the above action with the sigma model action of the form (\ref{A.1}),
the metric ${\tilde G}_{\Upsilon \Lambda}$ and the  tensor
field ${\tilde B}_{\Upsilon \Lambda}$ take the following forms
\begin{equation}\label{P.10}
{\tilde G}_{\Upsilon \Lambda} \;=\;\left( \begin{tabular}{cccc}
              $0$ & $1$ & $0$ & $0$\\
              $1$ & $0$ & $0$ & $0$ \\
              $0$ & $0$ & $0$ & $1$ \\
              $0$ & $0$ & $-1$ & $0$\\
                \end{tabular} \right),~~~~~~~~~~~~~~{\tilde B}_{\Upsilon \Lambda} \;=\;\left( \begin{tabular}{cccc}
              $0$ & $0$ & $0$ & $0$\\
              $0$ & $0$ & $0$ & $0$ \\
              $0$ & $0$ & $3{\tilde x}$ & $0$ \\
              $0$ & $0$ & $0$ & $0$\\
                \end{tabular} \right).
\end{equation}
Note that the background matrix of the dual model depends only on the bosonic field $\tilde x$. We see that in this example,
the super Poisson-Lie T-duality transforms the role of the fermionic field $\chi$ in the model (\ref{P.6}) to the bosonic field
$\tilde x$ on the dual model (\ref{P.10.1}).
Furthermore we observe that the metric is Ricci flat, i.e.,  $R=R_{_{\Upsilon \Lambda}}=0$.  As a WZW model, the original model
should be conformally invariant, but to  prove  the conformal invariance of
the dual model we  look at the
one-loop $\beta$-function equations.  Taking into account the dilaton field $\varphi$ for the sigma model
(\ref{A.1}), the one-loop $\beta$-function equations \cite{cosmology} are given by
$$
\beta_{_{\Upsilon \Lambda}}^{^{(G)}} = R_{_{\Upsilon \Lambda}}+\frac{1}{4} H_{_{\Upsilon \Delta \Xi}}
H^{^{\Xi \Delta }}_{\;\;~~\Lambda}+2\overrightarrow{\nabla}_\Upsilon
\overrightarrow{\nabla}_\Lambda \varphi=0,~~~\hspace{2cm}
$$
$$
~\beta_{_{ \Lambda \Delta}}^{^{(B)}}  =(-1)^\Upsilon  \overrightarrow{\nabla}^\Upsilon(
e^{-2\varphi} H_{_{\Upsilon\Lambda\Delta}}) =0,~~~\hspace{4cm}~
$$
\begin{equation}\label{P.11}
{{\beta}^{(\varphi)}}  = -R -\frac{1}{12} H_{_{\Upsilon\Lambda\Delta}}H^{^{\Delta\Lambda\Upsilon}}+4 \overrightarrow{\nabla}_\Upsilon \varphi \overrightarrow{\nabla}^\Upsilon \varphi  -4 \overrightarrow{\nabla}_\Upsilon \overrightarrow{\nabla}^\Upsilon \varphi=0,
\end{equation}
where
\begin{equation}\label{P.12}
H_{_{\Upsilon\Lambda\Delta}} \;=\; B_{_{\Upsilon\Lambda}}\frac{\overleftarrow{\partial}}{\partial
\Phi^{^{\Delta}}} +(-1)^{^{\Upsilon(\Lambda+\Delta)}}\; B_{_{\Lambda\Delta}} \frac{\overleftarrow{\partial}}{\partial
\Phi^{^{\Upsilon}}}+(-1)^{^{\Delta(\Upsilon+\Lambda)}}\; B_{_{\Delta\Upsilon}} \frac{\overleftarrow{\partial}}{\partial
\Phi^{^{\Lambda}}},
\end{equation}
is the torsion field.  We note that Eqs. (\ref{P.11})  are
equations of motion of the following effective action on
supermanifold \cite{cosmology}
\begin{equation}\label{P.12.1}
S_{eff}\;=\;\int d^{m, n}\Phi\; \sqrt{G}e^{-2\varphi} ( R+4{\overrightarrow{\nabla}_\Upsilon} {\varphi} \overrightarrow{\nabla}^\Upsilon {\varphi}
+\frac{1}{12} H_{_{\Upsilon\Lambda\Delta}}H^{^{\Delta\Lambda\Upsilon}}).
\end{equation}
After some calculations, one can find that the dilaton for the dual model is constant and
the only non-zero component of $H$ is $H_{233}=3$. Putting these
in (\ref{P.11}), one verifies that those equations are satisfied.
It is straightforward  to verify that $H_{_{\Upsilon\Lambda\Delta}}H^{^{\Delta\Lambda\Upsilon}}=0$, and
thus the effective action is vanished. Therefore, one can coclude
that the dual model is conformal invariant  for all loops. In
this way, {\it a pair of conformal sigma models related by super
Poisson-Lie T-duality is constructed by starting with the
$\left((C^3+A)\;,\;{C^3 \oplus {A}_{1,1}}_|.i\right)$  Drinfel'd
superdouble.}  Note that the original model (WZW model on the $(C^3+A)$) has also $R=R_{_{\Upsilon\Lambda}}=0$, $H_{144}=-1$ and $H_{_{\Upsilon\Lambda\Delta}}H^{^{\Delta\Lambda\Upsilon}}=0$,
such that the effective action for the original model is also vanished.
But this is not the whole story. Indeed, one can show that the dual action (\ref{P.10.1}) is the action of the WZW
model on the Lie supergroup $(C^3+A).i$ where its Lie superalgebra has the following non-zero (anti)commutation relations:
\begin{equation}\label{P.12.2}
[\hat{Z} , \hat{Q_+}]=\hat{Q_-},~~~~~~~\{\hat{Q_+} , \hat{Q_+}\}=-\hat{H}.
\end{equation}
Note that to construct the WZW
model (\ref{P.10.1}) on the Lie supergroup $(C^3+A).i$ we must choose the following convenient parametrization
\begin{equation}\label{P.12.3}
\hat{g}\;=\; e^{ (\frac{\hat{\chi}}{3}+\frac{\hat{x} \hat{\psi}}{2} ) \hat{Q_-}} ~ e^{\frac{-\hat{y}}{3} \hat{H}-  \hat{x}  \hat{Z}} ~ e^{ \hat{\psi}  \hat{ Q_+}}.~~~~~~~~~~~~~~~~~~~~~~~
\end{equation}
Thus, in this way, the super Poisson-Lie T-duality relates the WZW models to each other.

\section{Preservation of the N = (2,2) structure under super Poisson-Lie T-duality}

We know that the extended superconformal (especially N=2) current algebras are equivalent to the Manin triples \cite{parkh}.
The algebraic structure of the N = (2,2) supersymmetric WZW model is equivalent to the Manin triple (or Lie bialgebra) structure \cite{Ulf}.
Furthermore,  the study of the  N = (2,2) structure under the Poisson-Lie T-duality is  an interesting problem (see \cite{Eb}). Here in this paper according to the above
calculations, the super Poisson-Lie T-duality relates the WZW model on the Lie supergroup $(C^3+A)$  to the other WZW model on the Lie supergroup $(C^3+A).i$.
Thus, to construct the N = (2,2) WZW model on the Lie supergroup $(C^3+A)$, according to (\ref{L.9}) we have
\begin{equation}\label{N.1}
S_{N = (2,2) } (g) = \int d\xi^+ d\xi^-  d^{2}\theta~(-1)^{\Upsilon}
D_{+}\Phi^{\Upsilon} (G_{\Upsilon\Lambda}(\Phi)+B_{\Upsilon\Lambda}(\Phi)) D_{-}\Phi^{\Lambda},
\end{equation}
in which the superfields $\Phi^{\Upsilon}$ include all the bosonic coordinates $(y , x)$ and the fermionic ones $(\psi , \chi)$ (parameters of the Lie
supergroup $(C^3+A)$) where they are functions of the worldsheet coordinates $(\tau , \sigma)$ and Grassmannian  ones $(\theta_1 , \theta_2)$\footnote{Note that the action (\ref{N.1}) has been written in the N=1 formalism (see \cite{Ulf}).}, and
the background matrix is also given by (\ref{L.10}). By composing the invariance of the above action under the supersymmetry transformations
\begin{equation}\label{N.2}
{\delta^{1}}_{\hspace{-2mm}\epsilon}\;\Phi^{\Upsilon}=i(\epsilon^{+}Q_{+}+\epsilon^{-}Q_{-})\Phi^{\Upsilon},
\end{equation}
and
\begin{equation}\label{N.3}
{\delta^{2}}_{\hspace{-2mm}\epsilon}\;\Phi^{\Upsilon}=  \epsilon^{+}D_{+}\Phi^{\Lambda}{{J_+}^\Upsilon}_{\Lambda}(\Phi)
+\epsilon^{-}D_{-}\Phi^{\Lambda}{{J_-}^\Upsilon}_{\Lambda}(\Phi),
\end{equation}
where $Q_{\pm}$ and $D_{\pm}$ are
supersymmetry generators and superderivatives, respectively, the $\epsilon^{\pm}$ are the supersymmetry parameters and $J^{\Upsilon}_{\pm\Lambda} \in TG \bigotimes T^{\ast}G$, one can find that the $J^{\Upsilon}_{\pm\Lambda}$ must be complex structure on Lie supergroup $G$ where its generalized covariant derivative with respect to extended Christoffel symbols $(\Gamma^{\pm\Upsilon}_{~~\Lambda\Xi}=\Gamma^{\Upsilon}_{~\Lambda\Xi}\pm
(-1)^{\Omega} G^{\Upsilon\Omega}H_{\Omega\Lambda\Xi})$ must be zero \cite{Gat}, i.e., the background supermanifold $G$ must have super bi-Hermitian
structure. If we use the non-coordinate basis for Lie supergroup $G$ then we obtain the form of the algebraic
super bi-Hermitian structure as in the bosonic case \cite{Eb}, \cite{Sephid}.

We note that the $({\cal C}^3+{\cal A})$ Lie superalgebra is a
Drinfel'd super double as ${\cal D}=\left(({\cal A}_{1,1} +{\cal A}) ~,~{\cal I}_{(1|1)}
\right)$, so those equations automatically satisfy for the model (\ref{N.1}). Hence, the Lie supergroup
$({C}^3+{A})$ has a super bi-Hermitian structure or we have a N = (2,2) structure on the model (\ref{N.1}).
On the other hand, the dual model is also equivalent to the N=(2,2) WZW model on the Lie supergroup
$({C}^3+{A}).i$, because the $({\cal C}^3+{\cal A}).i$ Lie superalgebra is also a
Drinfel'd superdouble. In this way, the N=(2,2) structure is preserved under the super Poisson-Lie T-duality.
Note that this is the first example of this type.


\section{Conclusion and remarks}

We showed  that WZW model on the Lie supergroup $(C^3+A)$ contains
super Poisson-Lie symmetry when  the dual Lie supergroup is ${C^3
\oplus A_{1,1}}_|.i$. Then, we constructed  the mutually T-dual
sigma models on the Drinfel'd superdouble $\left((C^3+A)\;,\;{C^3
\oplus {A}_{1,1}}_|.i\right)$, and showed that the dual model is
also conformal invariant. It is quite interesting that the dual model is
equivalent to the WZW model on the Lie supergroup $(C^3+A).i$. Furthermore, using the fact that
the $({\cal C}^3+{\cal A}).i$ Lie superalgebra is a
Drinfel'd superdouble we saw that the N=(2,2) structure is preserved under the super Poisson-Lie T-duality.
Of course, there are some open problems related to our results. As an example, one can see that
WZW model on the Lie supergroup $(C^3+A).i$ (the dual model) has also super Poisson-Lie symmetry with
the dual Lie supergroup  $(2A_{1,1}+2A).i$ \cite{ER6}, and it seems this process can be continued.
In this way, we have a hierarchy of WZW models related by the super Poisson-Lie T-duality; such that it is different
from plurality, because the doubles are non-isomorphic. For exact discussion, one must first obtain and classify
all Lie superbialgebras of $({\cal C}^3+{\cal A})$, but this needs another challenge. Some of these problems are under investigation.

\bigskip
\bigskip

\noindent {\bf Acknowledgments:} This research was supported by a research
fund No. 401.231 from Azarbaijan Shahid Madani university.


\bigskip

\end{document}